\documentclass[manuscript=article, layout=twocolumn]{achemso}

\pdfoutput=1

\usepackage[version=3]{mhchem} 
\usepackage{color}
\usepackage{amsmath}

\author{Peter Loskill}
\author{Hendrik H\"ahl}
\author{Thomas Faidt}
\author{Samuel Grandthyll}
\author{Frank M\"uller}
\author{Karin Jacobs}
\email{k.jacobs@physik.uni-saarland.de}
\affiliation[Saarland University]
{Department of Experimental Physics, Saarland University, Saarbr\"ucken, 66041, Germany}

\title{Is adhesion superficial? Silicon wafers as a model system to study van der Waals interactions}

\begin{document}

\begin{abstract}
Adhesion is a key issue for researchers of various fields, it is therefore of uppermost importance to understand the parameters that are involved. Commonly, only surface parameters are employed to determine the adhesive forces between materials. Yet, van der Waals forces act not only between atoms in the vicinity of the surface, but also between atoms in the bulk material. In this review, we describe the principles of van der Waals interactions and outline experimental and theoretical studies investigating the influence of the subsurface material on adhesion. In addition, we present a collection of data indicating that silicon wafers with native oxide layers are a good model substrate to study van der Waals interactions with coated materials.

\end{abstract}

\section{Introduction}
\label{Intro}
Stimulating or preventing adhesion is a key issue for researchers of various fields.
To solve these problems, a comprehensive understanding of the prevailing adhesion mechanism is indispensable. Yet, not only various adhesions mechanisms exist, but also plenty of parameters that can affect adhesion: Nanoscale or microscale roughness~\cite{Fuller:1975td,Persson:2001ub}, static charges or the zeta-potential at the interface~\cite{Derjaguin:1941ww,Yoon:1992vg}, surface energies~\cite{Johnson:1971it, Good:1992tt}, and contact shapes~\cite{Spolenak:2005fi} are a few frequently-studied examples. All these parameters, however, have in common that they are describing the surface of a material. Hence, the question arises whether adhesion is really only 'superficial`. This question is of great importance since commonly used photoresists, coatings, adhesion promoters or other functionalized surface layers are often in the range of just a few nanometers. These dimensions are smaller than the range of interactions such as van der Waals (vdW) interactions. Hence, the material right underneath the surface might indeed have an effect on adhesion mediated by vdW interactions. In this paper, we review experimental and theoretical studies investigating the influence of subsurface material on vdW forces. Additionally, we provide a collection of experimental data highlighting the suitability of stratified substrates based on silicon wafers to study vdW interactions.

\section{Van der Waals interactions}

\begin{figure*}[t]
\centering
\includegraphics{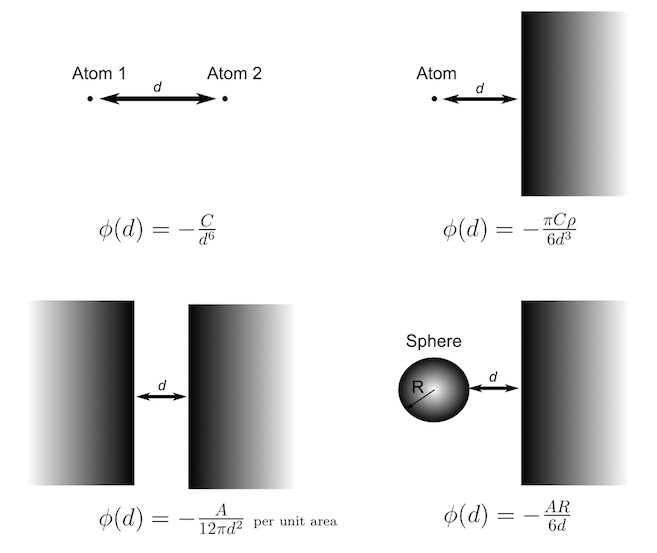}
\caption{VdW potentials $\phi(d)$ for different geometries. Adapted from~\cite{Israelachvili:1992vz}.}
\label{Fig:VDWEquations}
\end{figure*}

Already more than hundred years ago van der Waals introduced his theory of an attraction between neutral atoms in order to explain non-ideal gases~\cite{VanderWaals:1873up}. Later, three types of interactions were identified to contribute to the vdW interactions:

\begin{description}
\item[Keesom interactions] characterize dipol-dipol interactions of molecules that carry permanent dipoles~\cite{Israelachvili:1992vz}.
\item[Debye interactions] describe forces between a permanent dipole that induces a dipole moment in an otherwise unpolar molecule~\cite{Israelachvili:1992vz}.
\item[London interactions,] also called dispersion interactions, describe forces between instantaneously induced dipoles~\cite{London:1937uj}.
\end{description}
All three parts have in common that the interaction energy scales with $-\frac{1}{d^6}$. Hence, the vdW potentials $\phi$ for the interactions between two single atoms separated by a distance $d$ can be written as
\begin{equation}
\label{Eq:vdWAtoms}
\small \phi(d) = - C/d^6 
\end{equation}
Due to this scaling, vdW interactions are often considered as of short-range.

Hamaker, however, calculated energy-distance relations for macroscopic objects by pairwise summation over all atoms, continuing the work of Bradley and DeBoer~\cite{deBoer:1936ho,Bradley:1932tq,Hamaker:1937vm}. Depending on the geometry, different scaling laws apply (cf. Figure~\plainref{Fig:VDWEquations}). To account for the properties of the involved materials, Hamaker introduced a coefficient $A$, also called 'Hamaker constant`, which he defined to
\begin{equation}
\label{Eq:HamakerConstHamaker}
\small A = \pi^2 C \rho_1 \rho_2
\end{equation}
where $\rho_{i}$ are the number of atoms per unit volume of the two materials.
The controversial subject of Hamaker's theory was that he assumes a pairwise additivity of the vdW interactions, which is generally speaking not valid.
A few years later, Casimir used a completely different ansatz to calculate the force between two ideally conducting semi-infinite half-spaces in vacuum~\cite{Casimir:1948vq}.
On the basis of Planck's famous theory, he summed up the allowed electromagnetic modes between two conducting plates.
Lifshitz extended Casimir's idea and presented a theory for arbitrary materials, based on quantum field theory~\cite{Dzyaloshinskii:1961er}. In principle, although many studies differentiate between Lifshitz-vdW and Casimir interactions, Lifshitz and Casimir essentially described the same effect, but with different foci~\cite{French:2010hv, Parsegian:2006ux}. By treating the interacting objects as continuous media, Casimir's and Lifshitz' theories circumvent the question of pairwise additivity. Interestingly, Lifshitz' ansatz led to the same scaling laws as the classical Hamaker ansatz (cf. Figure~\plainref{Fig:VDWEquations}). Only the way the Hamaker constants are derived is different. Following Lifshitz' theory, they are calculated from the optical properties of the involved materials and can be approximated (see Appendix) by 
\begin{equation}
\small
\begin{split}
A_{12-32} \approx	 &\frac{3}{4} k_B T \left(\frac{\varepsilon_1-\varepsilon_2}{\varepsilon_1+\varepsilon_2}\right) \left(\frac{\varepsilon_3-\varepsilon_2}{\varepsilon_3+\varepsilon_2}\right)\\
&+ \frac{3 \hbar \omega_\textrm{\scriptsize e}}{8 \sqrt{2}} \frac{(n_1^2-n_2^2)(n_3^2-n_2^2)}{\sqrt{(n_1^2+n_2^2)(n_3^2+n_2^2)} }\\
&\quad  \cdot \frac{1}{\left[ \sqrt{(n_1^2+n_2^2)}+\sqrt{(n_3^2+n_2^2)}\right]}
\end{split}
\label{Eq:HamakerIsra}
\end{equation}
with the dielectric constants $\varepsilon_i$, the refractive indices in the visible regime $n_i$, and the main electronic absorption frequency $\omega_\textrm{\scriptsize e}$.

The works of Hamaker, Casimir and Lifshitz demonstrate that vdW-Casimir interactions can indeed be regarded as of long-range, since for mesoscopic and macroscopic objects, the absolute value of the exponent of the scaling law is decreased (the interaction between two semi-infinite half slabs scales with $-\frac{1}{d^2}$, for instance). Yet, the long-range character is restricted due to the finite speed of light~\cite{Casimir:1948bd}. This retardation effect increases the absolute value of the exponent of the scaling law by up to one (for $d\gg 20$\,nm). For separations smaller than 10\,nm, however, the retardation can usually be neglected~\cite{Tabor:1969wb, Israelachvili:1972ty}.

\section{Van der Waals interactions with coated substrates}
Using the equations given by the theories mentioned above, it is usually possible to predict the potentials for the interactions of two uniform objects.
Yet, many systems consist of coated substrates. However, VdW interactions act not only between atoms in the vicinity of the surface, but also between atoms in the bulk material.
Early experiments of Israelachvili und Tabor, showed that in the case of mica substrates covered with a monomolecular layer of stearic acids, both the acids and the mica contribute to the vdW interactions~\cite{Israelachvili:1972gn}. These experimental studies were in agreement with theoretical predictions of Langbein, who postulated that the interactions with the surface layer dominate for separations $d$ smaller than the layer thickness $D$ ($D\ll d$) and the interactions with the bulk material dominate in the opposing limit ($D\gg d$)~\cite{LANGBEIN:1971vb,LANGBEIN:1972vs}.
More recent studies showed that variations in the thickness of a surface layer induce differences in the vdW potentials and influence e.g.\ the stability of thin liquid coatings~\cite{Seemann:2001us, Seemann:2001ux, Jacobs:2008wb, Baeumchen:2010km}. These thin film dewetting studies moreover demonstrated quantitatively the impact of the subsurface composition to the effective interface potential~\cite{Vrij:1966gm}:  The impact was measured experimentally by determining the differences in the preferred wavelength of spinodally dewetting thin films with variable subsurface composition~\cite{Seemann:2001us, Herminghaus:1998ue}.

\subsection{Interactions in biological systems}

\begin{figure*}[t]
\centering
\includegraphics{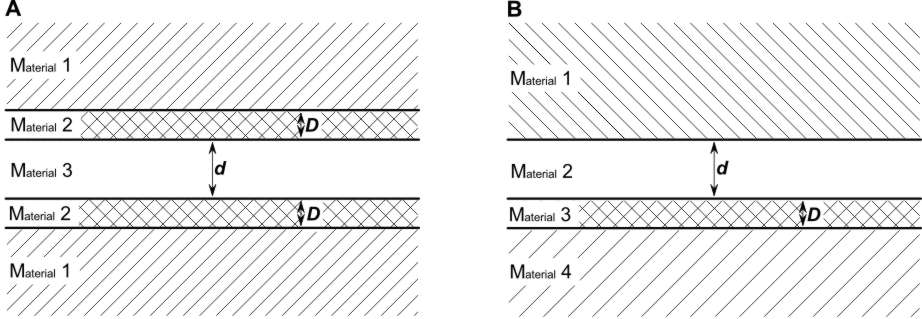}
\caption{Schemes of two multilayer configurations: A) symmetrical triple layers and B) interaction between a uniform material and a coated material.}
\label{Fig:LayerSystem}
\end{figure*}

VdW interactions also play a major role in biological systems~\cite{Leckband:2001tx}. Particularly non specific adhesion is governed by vdW interactions in conjunction with electric double layer interactions, usually described using the DLVO-theory~\cite{Derjaguin:1941ww, Verwey:1948uh} or extended DLVO-theory~\cite{VanOss:1989tf}. Many biological processes, such as the aggregation of proteins~\cite{LECKBAND:1994ta}, the unspecific adhesion of cells and bacteria~\cite{Nir:1977tb,Nir:1977ub,Ploux:2010jw}, the adherence of abalones~\cite{Lin:2009ir}, and the sticking of geckos~\cite{Autumn:2002hw, Autumn:2002tu, Huber:2005wx} are influenced and sometimes dominated by vdW interactions. As a logical consequence, these processes are also not `purely superficial', viz.\ not solely dependent on the properties of the surface. It could be shown that, when interacting with a coated substrate, proteins indeed sense both, the surface layer and the underlying material: Adsorption experiments on tailored silicon wafers with differences in the thickness of the oxide layer on top of the wafers revealed qualitatively different adsorption kinetics of multiple types of proteins~\cite{Quinn:2008dw, Schmitt:2010ke}. Based on Monte-Carlo simulations featuring surface processes such as surface mobility and conformational changes, the distinctions were invoked by the influence of the vdW-interactions on the time scale of these processes~\cite{Bellion:2008dl}. X-ray reflectivity experiments corroborated these findings~\cite{Hahl:WFuTT7jn}. Recent studies discovered that the influence of the subsurface material on adhesion is sensed by larger biological objects, too: The unspecific adhesion of bacteria from the \textit{Staphylococcus} genus is affected by the properties of the subsurface material, as could be shown by AFM force spectroscopy measurements~\cite{Loskill:2ew0dawZ}. Moreover, adhesion experiments with setal arrays of live geckos revealed that the adhesion force was significantly varied by a change in the substrates subsurface composition~\cite{Loskill:OBPQVRRF}.

\subsection{Theoretical description of multilayer systems}

For a comprehensive theoretical description of the vdW potentials for systems involving multilayer structures, not only the surface layer, but also the composition of the entire substrate has to be taken into account. Unfortunately, there is no general equation for the vdW interactions between arbitrary objects and layered substrates. In the following we therefore will focus on the non-retarded interactions between infinite planar interfaces.

Generally, a common way to find out unknown Hamaker constants is the use of combining rules (geometric mean)~\cite{VanOss:1988uq}. These relations, that are derived from the combining rules for surface energies, may be used to calculate `effective' Hamaker constants for a multilayer system. Yet, these relations break down whenever the Keesom and Debye part (the zero frequency terms) cannot be neglected~\cite{Israelachvili:1992vz}. Especially in multilayer systems, where multiple Hamaker constants are necessary, combining rules are not applicable. On the basis of the Lifshitz theory, the potentials for the interactions of symmetrical triple layer films (cf. Figure~\plainref{Fig:LayerSystem}\,A) were calculated by
\begin{equation}
\small 
\begin{split}
\phi_{\textrm{vdW}}(d) = -\frac{1}{12\pi}\cdot &\left(\frac{A_{23-23}}{d^2}\right. \\
&+ \left.2\frac{ A_{23-12}}{(d+D)^2}+ \frac{A_{12-12}}{(d+2D)^2} \right)
\end{split}
\label{Eq:SymTripleLayer}
\end{equation}
with $A_{ij-kl}$ the constants for the interactions of the two different interfaces~\cite{Ninham:1970vj, VAdrian:1973tf, Parsegian:1993wi}.
Using the same ansatz, the vdW potential of the interactions between a probe material and a substrate coated with a layer of thickness $D$ (cf. ~Figure~\plainref{Fig:LayerSystem}\,B) is given by~\cite{VAdrian:1973tf,Parsegian:2006ux}
\begin{equation}	
\label{Eq:CoatedLayer}
\small \phi_{\textrm{vdW}}(d) = -\frac{1}{12\pi}\cdot \left(\frac{A_{12-32}}{d^2} + \frac{A_{12-43}}{(d+D)^2} \right)\;.
\end{equation}

For larger separations $D \ll d$, however, Eq.~(\plainref{Eq:CoatedLayer}) is no longer valid. For the description of an experimental system with a variable $D$ (e.g.\ a thickness of a coating), we have previously chosen an alternative approximation~\cite{Seemann:2001us, Jacobs:2008wb}: By assuming a scaling of the interaction with $ \frac{C_a}{d^2}+\frac{C_b}{(d+D)^2}$ and a continuous transition between the boundary cases $D \ll d$ and $D \gg d$, we gained 
\begin{equation}
\label{Eq:SeeCoatedLayer}
\small \phi(d) = -\frac{1}{12\pi}\cdot \left(\frac{A_{12-32}}{d^2} + \frac{A_{12-42}-A_{12-32}}{(d+D)^2} \right)\;,
\end{equation}
where $A_{12-42}$ stands for the interaction of material 1 via medium 2 with material 4 in the case that $D = 0$, viz.\ medium 3 is nonexistent.

\section{Silicon wafers as a model system}

\begin{figure*}[t]
\centering
\includegraphics{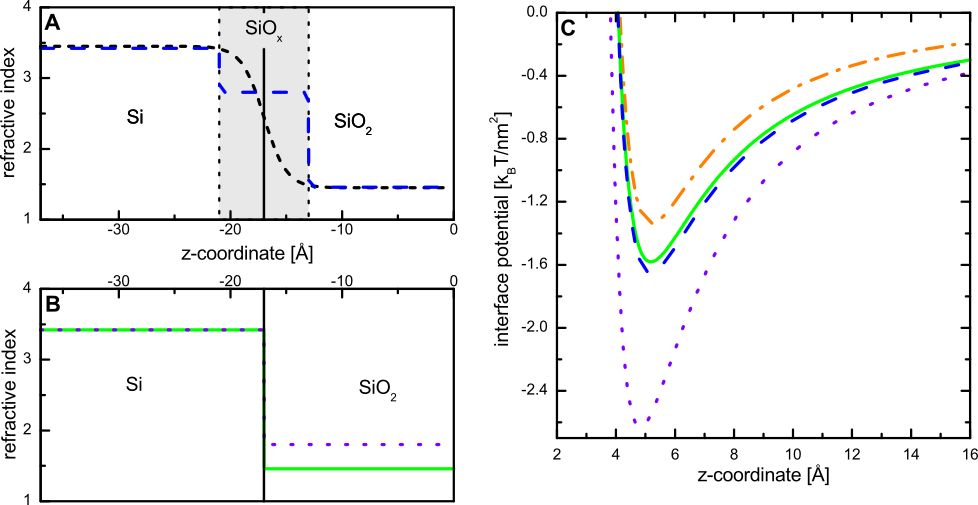}
\caption{Properties of thin oxide layers: A) Theoretically expected transition of the refractive index (short dashes) and a proposed double layer oxide model (long dashes). B) Single layer oxide models whereby i) the oxide layer has i) SiO$_2$ bulk-like properties (solid line) or ii) properties expected by optical studies with limited resolution (dotted line). C) Calculated potentials of the van der Waals interactions between a semi-infinite half-space of polystyrene ($n_\textrm{vis}= 1.585$, $ \epsilon_0 = 2.6$) and a silicon wafer with a thin oxide layer for the three model configurations (cf.\ A and B) and a bulk silicon dioxide substrate ($n_\textrm{\scriptsize vis}= 1.46$, $ \epsilon_0 = 3.9$ \cite{Sze:2006uha}) as a comparison (dash-dotted line). To include the short-range repulsive interaction a second term $\frac{C}{d^8}$ (maintaining the difference in exponents of the Lennard-Jones potential) was added. The parameter $C$ was kept constant for all configurations, since the surface is essentially the same.}
\label{Fig:ThinSiOFilms}
\end{figure*}

Most of the studies mentioned above used silicon wafers as a model system to study vdW interactions with coated materials. As described earlier, these interactions are essentially dependent on the optical properties of the involved materials (Eq.~(\plainref{Eq:HamakerIsra})). The properties of  silicon oxide films, however, have been discussed controversially for decades. In the following, the term ``silicon oxide'' stands for SiO$_x$. The use of SiO$_2$ bulk values for oxide films thicker than 100\,nm is generally accepted. Yet, the validity of these values for thin film has been questioned numerous times and especially the optical properties, such as the refractive index, are disputed. In general, thin silicon oxide layers may be described by two different models:
\begin{description}
\item[Single layer model:] By assuming a sharp transition between the bulk Si and the oxide, the latter can be described by a single layer (cf. Figure~\plainref{Fig:ThinSiOFilms}\,A). In this case, the oxide layer may have the same refractive index as bulk SiO$_2$ or an increased refractive index.
\item[Double layer model:] Since a sharp transition between the materials is highly unphysical, a continuous transition or an interface roughness is very likely. As the thickness of this transition region is of the same order of magnitude as the oxide layer thickness, a double layer configuration is a obvious approximation (cf. Figure~\plainref{Fig:ThinSiOFilms}\,B).
\end{description}
The van der Waals interactions sensed by an probe object should differ strongly for these different model configurations (cf.Figure~\plainref{Fig:ThinSiOFilms}\,C).

\begin{table*}[t]
\centering
\begin{tabular}{ccccccc}
			oxide layer &$d$ {\scriptsize [nm]}& $rms$ {\scriptsize [nm]}& $\varTheta_{\text{adv}}$ {\scriptsize [$^\circ$]}	& $ \varTheta{_\text{rec}}${\scriptsize  [$^\circ$]}& $\gamma$  {\scriptsize [mJ/m$^{2}$]} 	\\\hline
native & 1.7(3)	& 0.13(3)		& 5(2)		& {\scriptsize compl.\ wetting}	& 63(1)		\\
thick & 150(1)	& 0.09(2)		& 7(2)	& {\scriptsize compl.\ wetting}		& 64(1)	\\\hline
\end{tabular}
\caption{Surface properties of a native oxide layer and of a thermally grown thick oxide layer as a comparison: thickness ($d$), root mean square ($rms$) roughness (by (1$\mu$m)$^2$ AFM scan), advancing (adv) and receding (rec) water contact angle, and surface energy $ \gamma$ (obtained from contact angle measurements of three different liquids~\cite{Mykhaylyk:2003cw}).}
\label{Tab:SurfaceProperties}
\end{table*}

By applying the single layer model\bibnote{Jellison indeed assumed a transition layer. Yet, at some point (for thin films) he started to neglect this layer, whereby the increase in refractive index began exactly around this point.}, Jellison observed an increase in the refractive index of the whole oxide layer~\cite{JellisonJr:1991tf}. Using spectroscopic polarization modulation ellipsometry, the refractive index of very thin layers was determined to $1.5-1.8$ (at $\lambda = 800$\,nm). These findings were later on confirmed by other studies~\cite{Wang:2000th,Hebert:1996vt}.

Experimental support for the double layer model, viz.\ the observation of an interfacial transition layer, is also given by previous studies. Experimental studies noticed a thin ($\approx 6-7$\,\AA) region of atomically mixed Si and O with an refractive index of $n \approx 2.8-3.2$ (at 546.1\,nm)~\cite{TAFT:1979vt, ASPNES:1979wq}. High-resolution core-level and XPS spectroscopy~\cite{Hollinger:1984cm, GRUNTHANER:1979tj} also confirmed that ``the interface is not abrupt, as evidenced by the high density of intermediate-oxidation states (about two monolayers of Si) and by their nonideal distribution''~\cite{Himpsel:1988ub}. These findings were matched by predictions of theoretical models~\cite{Ohdomari:1987hr, Giustino:2005hn}.

The results of theses studies, however, are not contradictory, but arise from the different methods applied. A problem of optical reflectivity methods, such as ellipsometry, is that they are not able to determine the density and the thickness of thin films ($\leq 5\,$nm) independently (not to mention to distinguish between two of such films). Thus, the usage of the single layer model for these methods is the only possible way. Yet, for thinner silicon oxide films, the transition layer fraction of the total oxide layer is increased resulting in an observation of a higher overall refractive index.

Another limitation of all of the mentioned studies is, that they are performed on silicon wafers with thermally grown silicon oxide layers. Especially for thin ($\leq 10$\,nm) films, the process parameters of the artificial growing process can have a significant influence on the density and the optical properties of the silicon oxide (decreased vs.\ increased refractive index)~\cite{Irene:1982vi, Cai:2010ka}. Yet, already without any pretreatment, silicon wafers are covered with a native oxide layer of 1.5\,nm to 2\,nm thickness. Since only limited data is available for the optical properties of native oxide layers, we present a brief summary of the properties of these layers containing previously unpublished data.

\subsection{Properties of thin native oxide layers}

The increase in refractive index to values up to $n=1.8$, as predicted by several studies~\cite{JellisonJr:1991tf,Wang:2000th,Hebert:1996vt}, is highly unlikely for native oxide layers as polymer dewetting studies have shown via an indirect way~\cite{Seemann:2001us}: Thin liquid polystyrene (PS) films prepared on Si wafers with native oxide layers ($D_{\text{SiO}}= 2.4\,$nm) were unstable and dewetted spinodally. Since this process is driven by the minimization of the free energy determined by the vdW potential~\cite{Geoghegan:2003td,Israelachvili:1992vz} the refractive indexes of the oxide cannot be higher than the one of PS ($\approx 1.59$), as shown by Eq.~(\plainref{Eq:HamakerIsra}).

\begin{figure}[t]
\centering
\includegraphics{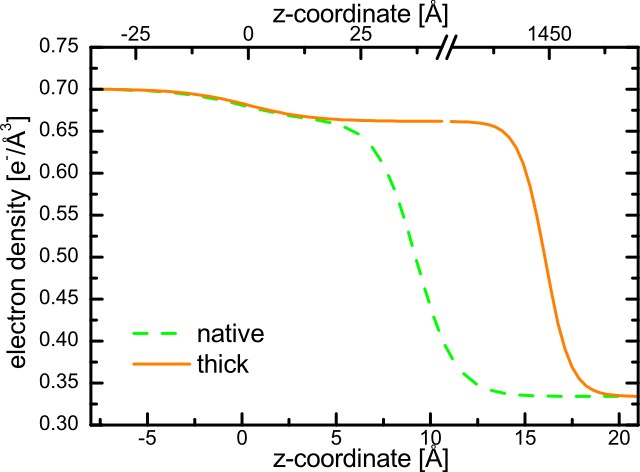}
\caption{Electron density profiles of a native (dashed line) and a thick oxide layer (solid line) determined by high energy X-ray reflectivity. The z-coordinate has been set to zero at the center of the transition between Si and its oxide.}
\label{Fig:XrayThinSiO}
\end{figure}

\begin{figure*}[t]
\centering
\includegraphics{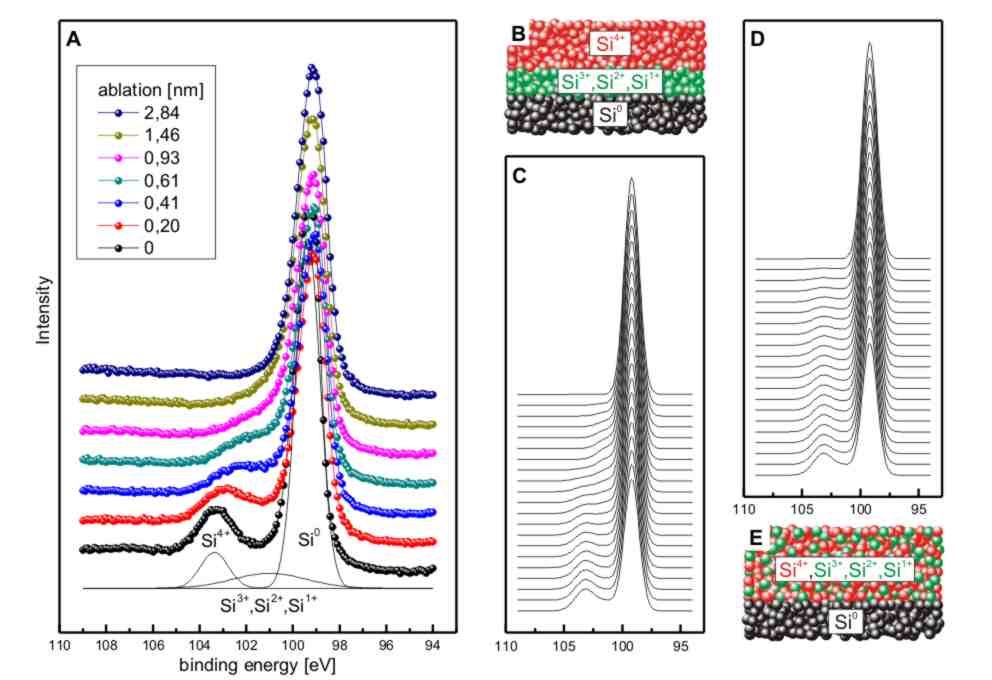}
\caption{A) XPS core level spectra for Si-2p (Al-Ka radiation, $\hbar \omega = 1486.6$\,eV, normal emission) for different stages of surface ablation by Ar ion etching with asymmetric decrease of intensity contributions from oxide species. B) Double layer oxide model with Si$^{0}$ - (Si$^{1+}$, Si$^{2+}$, Si$^{3+}$) - Si$^{4+}$ stacking (from bulk to surface). C) Simulation of spectra for the double layer model from B) with the same asymmetric decrease of oxide intensities as observed in experiment. D) Simulation of spectra for the single layer model depicted in E) with uniform decrease of oxide intensities. E) Single layer model with a homogeneous (Si$^{1+}$, Si$^{2+}$, Si$^{3+}$, Si$^{4+}$) surface layer.}
\label{Fig:XPSThinSiO}
\end{figure*}

A more direct ansatz is to compare native oxide layers to thick oxide layers in terms of the material properties, such as surface roughness, surface chemistry, homogeneity, electron density and stoichiometry: The surface characterization via atomic force microscopy (AFM) and contact angle (CA) measurements shows that - within the experimental error - the roughness and chemical homogeneity of the surface of a native oxide layer on Si wafers does not differ from the properties of a thick SiO$_2$ layer (cf. Table~\plainref{Tab:SurfaceProperties}). The analysis of high energy X-ray reflectivity measurements on native and thick oxide layers leads to electron densities that are again similar (Figure~\plainref{Fig:XrayThinSiO}). Especially the topmost part ($0-5$\,\AA) of the native oxide layer resembles the density of the thicker layer.

The stoichiometry of the native oxide layer was studied by X-ray photoelectron spectroscopy (XPS) combined with Ar ion etching in order to reveal the depth profiles for different oxidation states/valencies Si$^{k+}$ ($k = 0,\dots ,4$). Figure~\plainref{Fig:XPSThinSiO}\,A shows the Si-2p spectra recorded in normal emission mode (take-off angle 0$^\circ$ along the surface normal) with two components representing Si$^0$ and Si$^{4+}$ at lower and higher binding energy, respectively. Contributions from Si$^{1+}$, Si$^{2+}$ and Si$^{3+}$ could not be resolved, but have to be treated as a third peak in the background between the Si$^0$ and Si$^{4+}$ signals.
For stepwise ablation of the surface, the intensity of the Si$^{4+}$ peak decreases asymmetrically, i.e., it is shifted towards the Si$^0$ signal, forming a shoulder in the intermediate state before vanishing (for calibration of ablation see, e.g., Ref.~\cite{Mueller:2010bca}).
For ablation of about 1.5\,nm, contributions from oxide species can no longer be observed, which is in accordance with the thickness of the native oxide measured by other methods (Figure~\plainref{Fig:XrayThinSiO} and Table~\plainref{Tab:SurfaceProperties}).
The asymmetry in the Si oxide related part of the spectra is characteristic for intensity contributions from Si$^{4+}$ and (Si$^{1+}$, Si$^{2+}$, Si$^{3+}$) when distributed in a double layer model as depicted in Figure~\plainref{Fig:XPSThinSiO}\,B. For this type of stacking, the simulation of intensity distributions within the Si-2p spectra in Figure~\plainref{Fig:XPSThinSiO}\,C shows the same characteristics, namely the asymmetric decrease of the intensity from the oxide species as the experimental data in Figure~\plainref{Fig:XPSThinSiO}\,A. A similar asymmetry was reported in previous studies on thermally grown oxide layers~\cite{GRUNTHANER:1979tj}. For comparison, Figure~\plainref{Fig:XPSThinSiO}\,E shows a second model with a homogeneous distribution of the oxidation states within the oxide layer. For this scenario, the simulation in Figure~\plainref{Fig:XPSThinSiO}\,D predicts a uniform disappearance of the oxide contributions, in contrast to the experimental observations in Figure~\plainref{Fig:XPSThinSiO}\,A. 

In summary, these results show that native silicon oxide layers can be approximated by a double layer system as depicted in Figure~\plainref{Fig:ThinSiOFilms}\,A and Figure~\plainref{Fig:XPSThinSiO}\,B, where adjacent to the bulk Si a transition layer and then a bulk-like SiO$_2$ layer follows. Since all characterized material properties of the SiO$_2$ part are similar to the properties of thick oxide layers, there are no hints to assume different optical properties. The transition layer, however, will most likely display increased polarizability~\cite{TAFT:1979vt, ASPNES:1979wq,Giustino:2005hn}. That means, differences in the vdW potentials wafers featuring native and thick oxide layers arouse indeed from subsurface contributions, since the uppermost material is the same, namely SiO$_2$. Hence, silicon wafers with native oxide layers are indeed a good model substrate to study van der Waals interactions with coated materials. 

\section{Appendix}
\subsection{Calculation of Hamaker constants}
\label{HamakerConstants}
Based on the Lifshitz theory, the Hamaker constant can be calculated by\bibnote{The prime denotes that the zero term has to be multiplied with $\frac{1}{2}$.}
\begin{equation}
\small A_{12-32} = \frac{3}{2} k T \sideset{}{^\prime}\sum_{n=0}^\infty \Delta_{1,2}(i \xi_n) \Delta_{3,2}(i \xi_n)
\label{Eq:HamakerLifshitz}
\end{equation}
with 
\begin{equation}
\small \Delta_{a,b}(i \xi) = \frac{\varepsilon_a (i \xi) - \varepsilon_b (i \xi)}{\varepsilon_a (i \xi) + \varepsilon_b (i \xi)}
\end{equation}
whereby $\varepsilon_a (i \xi)$ are the values of the dielectric function of material $a$ at the imaginary (Matsubara) frequencies
\begin{equation}
\label{Eq:Matsubara}
\small \xi_n = n \cdot 2\pi k_B T/h
\end{equation}

Using the Ninham-Parsegian approximation it is possible to obtain the $\varepsilon (i \xi)$ from the adsorption spectrum~\cite{Parsegian:1969ub,Ninham:1970vj}, more precisely the relative strengths and the frequencies of the peaks, by
\begin{equation}
\small \varepsilon (i \xi) = 1 + \sum_{j=0}^N \frac{C_j}{1+(\xi/\omega_j)^2}
\label{Eq:Epsilon}
\end{equation}
with
\begin{equation}
\small C_j = \frac{2 f_j}{\pi \omega_j}
\end{equation}
where $f_j$ is the strength of an oscillator, $\omega_j$ its relaxation frequency, and N is the total number of oscillators.
For dielectric materials Eq.~(\plainref{Eq:Epsilon}) reduces to~\cite{MAHANTY:1976wm, Hough:1980go}
\begin{equation}
\small \varepsilon (i \xi) 
\approx 1 + \frac{\varepsilon(0)-\varepsilon(\omega_\textrm{\scriptsize vis})}{1+(\xi/\omega_\textrm{\scriptsize rot})^2}+ \frac{\varepsilon(\omega_\textrm{\scriptsize vis})-1}{1+(\xi/\omega_\textrm{e})^2}
\label{Eq:ReducedDielectric}
\end{equation}
with $\omega_\textrm{\scriptsize rot}$ the molecular rotational relaxation frequency (typically in the IR regime), $\omega_\textrm{\scriptsize e}$ the main electronic absorption frequency (typically $\approx 3 \cdot 10^{15}$~s$^{-1}$, in the UV regime), and $\varepsilon(\omega_\textrm{\scriptsize vis}) = n_\textrm{\scriptsize vis}^2$ the refractive index in the visible regime. As usually $\xi_1 >> \omega_\textrm{\scriptsize rot}$ (cf. Eq.~(\plainref{Eq:Matsubara})) the first term in Eq.~(\plainref{Eq:ReducedDielectric}) can be neglected and Eq.~(\plainref{Eq:HamakerLifshitz}) may be approximated by Eq.~(\plainref{Eq:HamakerIsra}) if the adsorption frequencies of the three materials are similar~\cite{Israelachvili:1992vz}.

\providecommand*\mcitethebibliography{\thebibliography}
\csname @ifundefined\endcsname{endmcitethebibliography}
  {\let\endmcitethebibliography\endthebibliography}{}


\begin{mcitethebibliography}{72}
\bibitem[Not()]{Note-2}
The prime denotes that the zero term has to be multiplied with
  $\frac{1}{2}$.\relax
\mciteBstWouldAddEndPunctfalse
\mciteSetBstMidEndSepPunct{\mcitedefaultmidpunct}
{}{\mcitedefaultseppunct}\relax
\EndOfBibitem
\end{mcitethebibliography}
\end{document}